\documentclass[aps,prb,twocolumn, superscriptaddress]{revtex4-1}
\usepackage[latin2]{inputenc}
\usepackage{float}
\usepackage{graphicx}
\usepackage{hyperref}
\hypersetup{colorlinks=false}
\usepackage{amsmath}
\usepackage{xcolor}
\usepackage[mathscr]{eucal}

\begin{document}

\title{Voltage-controlled binary conductance switching in gold--4,4'-bipyridine--gold single-molecule nanowires}

\author{G.~Mezei}
\thanks{These authors contributed equally to this work.}
\affiliation{Department of Physics, Budapest University of Technology and Economics,\\ 1111 Budapest, Budafoki ut 8., Hungary}
\affiliation{MTA-BME Condensed Matter Research Group, Budafoki ut 8, 1111 Budapest, Hungary}

\author{Z.~Balogh}
\thanks{These authors contributed equally to this work.}
\affiliation{Department of Physics, Budapest University of Technology and Economics,\\ 1111 Budapest, Budafoki ut 8., Hungary}
\affiliation{MTA-BME Condensed Matter Research Group, Budafoki ut 8, 1111 Budapest, Hungary}

\author{A.~Magyarkuti}
\affiliation{Department of Physics, Budapest University of Technology and Economics,\\ 1111 Budapest, Budafoki ut 8., Hungary}
\affiliation{MTA-BME Condensed Matter Research Group, Budafoki ut 8, 1111 Budapest, Hungary}

\author{A.~Halbritter}
\affiliation{Department of Physics, Budapest University of Technology and Economics,\\ 1111 Budapest, Budafoki ut 8., Hungary}
\affiliation{MTA-BME Condensed Matter Research Group, Budafoki ut 8, 1111 Budapest, Hungary}

\date{\today}

\begin{abstract}
We investigate gold--4,4'-bipyridine--gold single-molecule junctions with the mechanically controllable break junction technique at cryogenic temperature ($T=4.2\,\text{K}$). We observe bistable probabilistic conductance switching between the two molecular binding configurations, influenced both by the mechanical actuation, and the applied voltage. We demonstrate that the relative dominance of the two conductance states is tunable by the electrode displacement, whereas the voltage manipulation induces an exponential speedup of both switching times. The detailed investigation of the voltage-tunable switching rates provides an insight into the possible switching mechanisms.
\end{abstract}

\maketitle

\footnotetext{\textit{$^{\ddag}$~These authors contributed equally to this work.}}

%\section{Introduction}
The field of single-molecule electronics\cite{molelectr} envisions the exploitation of single molecule interconnects between conducting electrodes as functional devices. A key component towards this goal is the well-controlled manipulation of the conductance states by external parameters. \cite{Poulsen_review,switch_review1,switch_review2,switch_review3} Single-molecule conductance switching was already induced by mechanical actuation, \cite{BPY1, Su2015} light irradiation, \cite{Roldan2013,doi:10.1002/advs.201500017,Tam2011,Kim2012,Meng2019} or by the voltage applied on the junction. \cite{Collier391,voltage_switch1,voltage_switch2,voltage_switch3,voltage_switch4,voltage_switch5,voltage_switch6} In the latter case, usually complex molecules are applied, where either a charged or polarizable segment of the molecule is actuated by the electric field or the charge state of the molecule is manipulated. Here, we demonstrate voltage-tunable probabilistic conductance switching with a simple benchmark molecule, 4,4'-bipyridine (BP).

Previous measurements on Au-BP-Au single-molecule structures have shown, that the BP molecule can attach to gold electrodes in two different binding geometries, resulting in double-step molecular plateaus on the conductance versus electrode separation traces\cite{BPY1,BPY2,force6, Borges2016, magyarkuti17, Isshiki2018, magyarkuti2020} after the rupture of a single-molecule gold junction with $G\approx2e^2/h=1\,$G$_0$ conductance. It is suggested, that at a smaller electrode separation, the molecule binds on the side of the metallic junction, such that both the nitrogen linker and the aromatic ring are electrically coupled to the metal electrode ($G\approx 10^{-3}\;$G$_{0}$). Upon increasing the gap, the molecule slides to the apex and only the linkers couple to the electrodes, yielding a decreased junction conductance ($G\approx 3 \cdot 10^{-4}\;$G$_{0}$). \cite{BPY1,BPY2, force6} In the following, we refer to these two binding geometries as \emph{HighG} and \emph{LowG} configurations, which are also reflected by two clear peaks on the conductance histograms. A mechanically-controlled binary conductance switching between these two configurations was also reported,\cite{BPY1} however, the voltage-controlled manipulation of the conductance states was hindered by the limited junction stability in the room temperature break junction environment.  

Here we utilize a cryogenic temperature ($T=4.2\;$K) mechanically controllable break junction (MCBJ) setup to investigate the voltage controlled binary conductance switching in Au--BP--Au single-molecule junctions. The superior low-temperature stability enables us to map the response of the junction to the manipulation of the external parameters in detail. 
We observe bistable switching between the two molecular configurations, influenced both by the mechanical actuation, and the applied voltage. We find that at low bias voltage, the electrode separation determines the preferred junction configuration, similarly to prior room temperature measurements.\cite{BPY1} However, as the voltage is increased, a probabilistic switching is induced. We demonstrate that the relative dominance of the two configurations can be tuned through adjusting the electrode separation, while the switching rates are controlled by the applied voltage. This behaviour resembles the characteristics of probabilistic bit devices.\cite{Camsari2019, Borders2019} The detailed investigation of the voltage-tunable switching rates provides an insight into the possible switching mechanisms.

\section{Results and Discussion}
\begin{figure}
    \centering
    \includegraphics[width=\columnwidth]{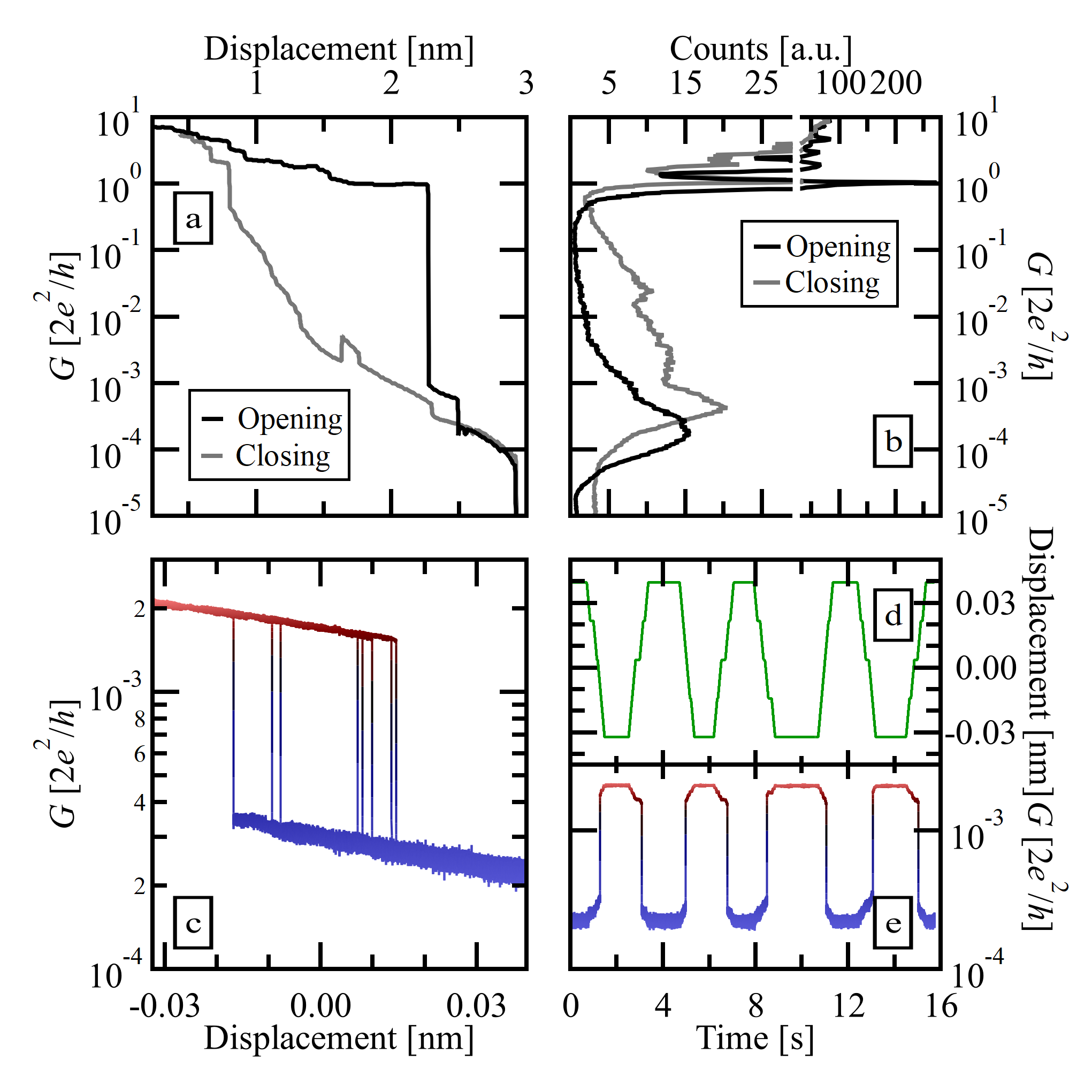}
    \caption{\textbf{a)} Sample opening (black) and closing (grey) trace pairs at $T=4.2\;$K temperature.  
    \textbf{b)} Opening (black) and closing (grey) conductance histograms at $T=4.2\;$K. \textbf{c)} A hysteretic conductance switching is observed during repeated displacement cycles. \textbf{d-e)} The time dependence of the measurement in panel c): the periodical displacement  modulation (d) yields a repeated conductance switching between the HighG (red) and LowG (blue) configurations (e). Utilizing the atomic chain pulling mechanism of gold \cite{yanson98} we have used the interpeak distance in the length distribution of the $1\,$G$_0$ plateaus to calibrate the displacement axis.\cite{plateaulength}}
    \label{fig:fig1}
\end{figure}

We performed measurements using a cryogenic MCBJ setup with high purity gold wires. After cooling down the setup, a histogram is recorded to ascertain the cleanliness of the junction, then molecules are dosed using an in-situ evaporation technique.\cite{magyarkuti17,magyarkuti2020} We show an example pair of conductance traces in Fig.~\ref{fig:fig1}a, black corresponds to the opening while grey to the closing trace, both exhibiting molecular plateaus. While the room temperature conductance histograms acquired by the same setup\cite{magyarkuti17} reproduce the common double-peak structure,\cite{BPY1,BPY2, Borges2016} the histogram for our low-temperature measurement shows a significant difference (black curve on Fig.~\ref{fig:fig1}b), exhibiting an enhanced peak at the LowG interval and only a slight shoulder in the HighG region. In our previous work,\cite{magyarkuti2020} we have shown that this phenomenon is related to the low-temperature single-atomic chain pulling mechanism typical for gold.\cite{yanson98,ohnishi98,plateaulength} The probable atomic chain formation process in the stable low-temperature environment yields such wide gaps after the chain rupture that are unable to accommodate the HighG molecular configuration.

Compared to the low-temperature opening histogram, the peaks in the closing histogram (grey curve in Fig.~\ref{fig:fig1}b) are shifted towards higher values. Furthermore, the HighG peak becomes more pronounced, although the LowG peak remains dominant. We explain this with the strain acting on the junction in the opening process. This strain stretches the molecule in the contact, lowering the measured conductance of the junction. However, in the closing process, the electrodes and the molecule are relaxed, no strain is present, which leads to a somewhat higher conductance.

Next, we investigate how changing the electrode separation affects the molecular binding configuration.  In their prior study  Quek et al. reported mechanically-controlled binary conductance switching between the two binding configurations.\cite{BPY1} This was realised by stopping the opening process and applying a short ($\sim 60\,\text{ms}$) periodic signal to the piezo actuator to mechanically perturb the junction by repeatedly compressing and stretching it with $0.2\;$nm amplitude. When a molecule was caught in the junction during the period of the mechanical perturbation, the conductance indicated periodic switching between the two binding configurations. In  Figs.~\ref{fig:fig1}c,d,e, we reproduce this mechanical switching at 2 orders of magnitude longer time-scales, utilizing the enhanced mechanical stability of the junction in the low-temperature environment.

\begin{figure}
    \centering
    \includegraphics[width=\columnwidth]{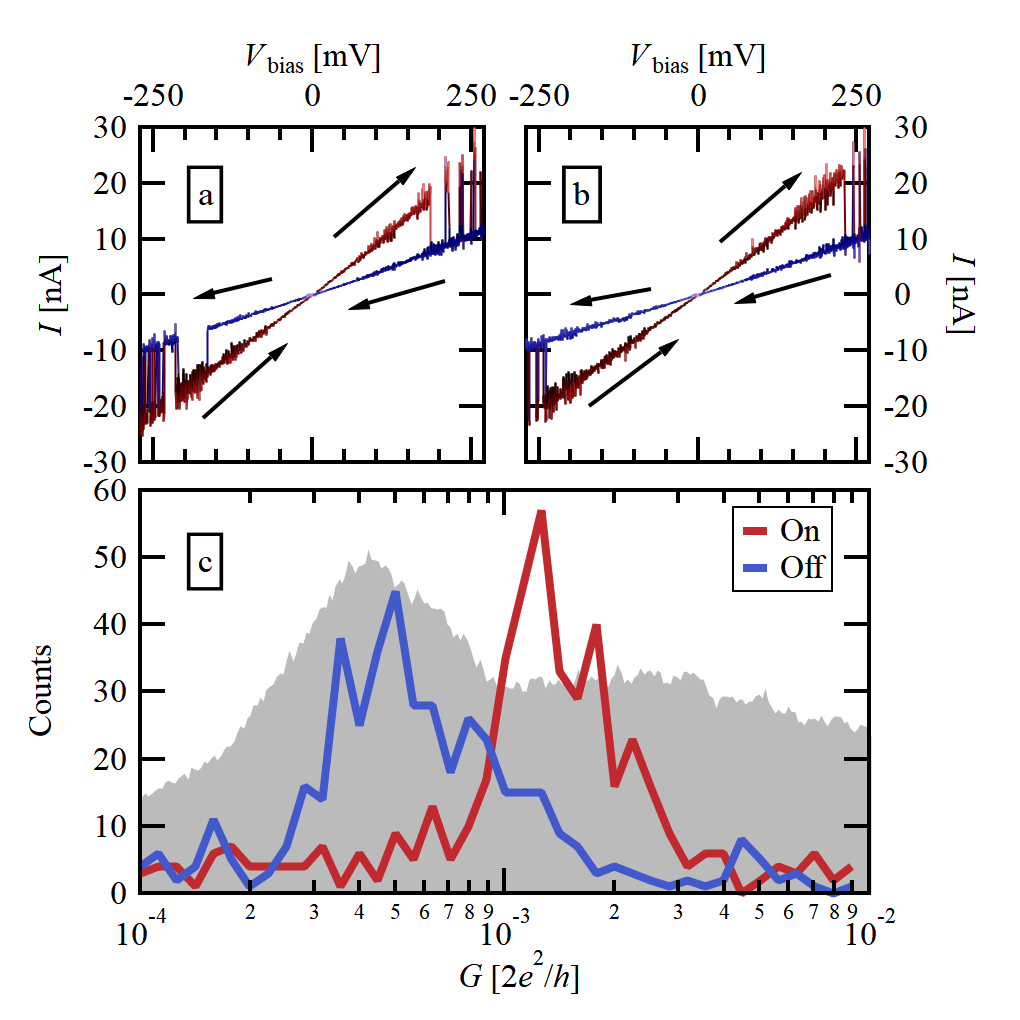}
    \caption{\textbf{a-b)} Sample $I(V)$ curves measured at $T=4.2\,\text{K}$. The arrows indicate the direction of the  voltage-ramp. \textbf{c)} The distribution of the conductance values determined from the slope of 465 I(V) curves for the on (red) and off (blue) states. The grey shaded area serves as a comparison basis reproducing the closing conductance histogram from Fig.~\ref{fig:fig1}b.
    }
    \label{fig:fig3}
\end{figure}

We continue by investigating the current-voltage characteristics of BP single-molecule junctions. Typical $I(V)$ curves (Fig.~\ref{fig:fig3}a,b) show that at low bias voltage, the present state of the junction is stable. However, when the applied voltage is increased, the junction exhibits binary probabilistic switching between two distinct conductance states. Upon further increasing the voltage, the switching rates also increase. Fig.~\ref{fig:fig3}c shows the distribution of the high and low conductance states (i.e. the \emph{on} and \emph{off} conductances) deduced from a statistical amount of switching $I(V)$ curves. As a reference, the closing conductance histogram is reproduced from Fig.~\ref{fig:fig1}b as a grey area graph. The similar positions of the peaks in the closing histogram and the distributions of the on and off conductance states suggests that the molecular junction is switching between the HighG and LowG binding configurations, as observed during the closing of the junction. This result is consistent with our previous argument that the difference between the opening and closing histograms is caused by the strain acting on the junction in the opening process. During these $I(V)$ measurements, the junction has time to relax, so that the strain acting on the molecule is reduced, leading to similar HighG and LowG conductance values as seen in the closing conductance histogram. These findings demonstrate that besides mechanical, we can also achieve electric switching between the two configurations of the BP molecule.

\begin{figure}
    \centering
    \includegraphics[width=\columnwidth]{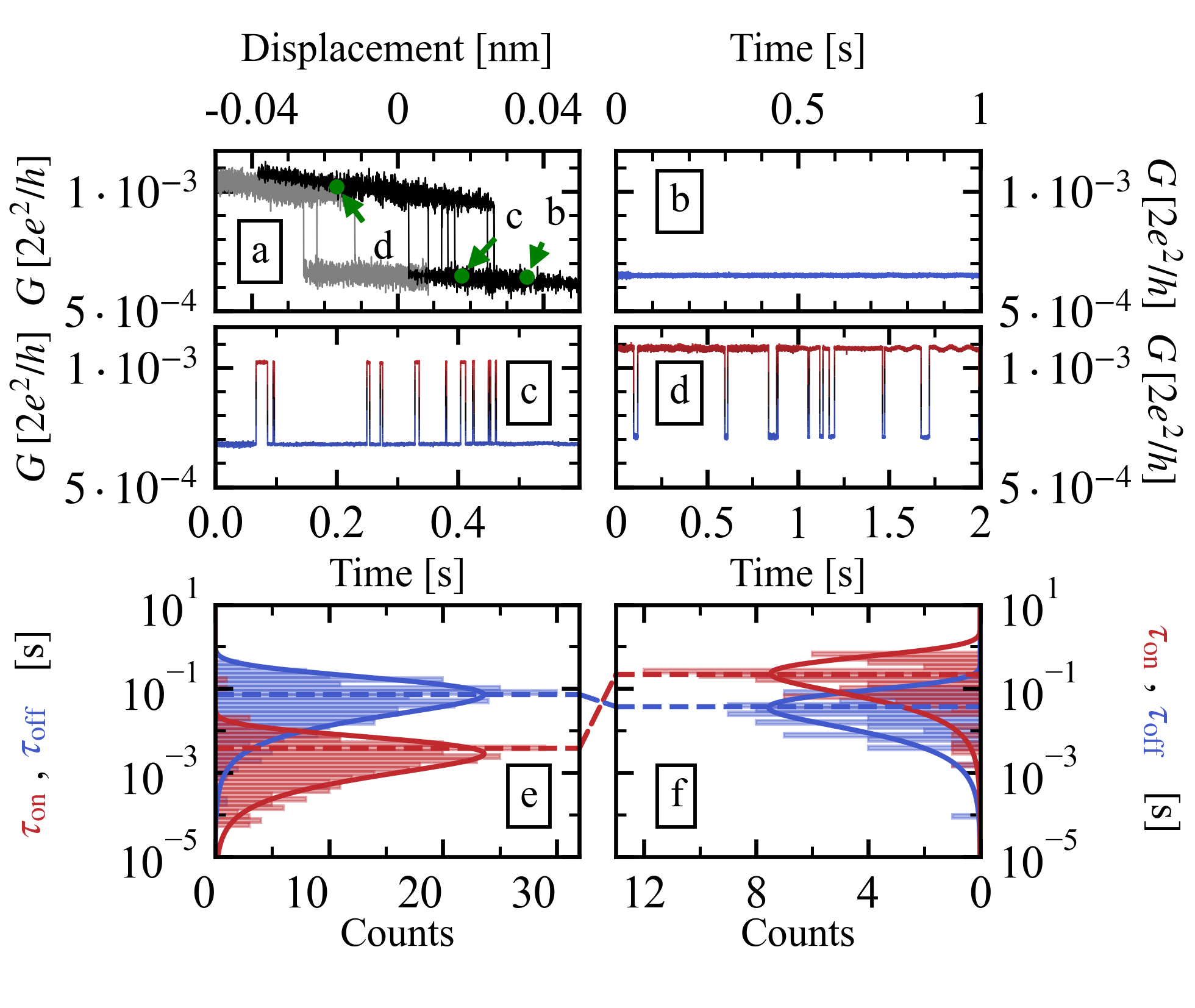}
    \caption{\textbf{a)} Hysteretic conductance jumps between the HighG  and LowG states along piezo ramp cycles recorded at low bias ($V=30\,$mV). The grey hysteresis was recorded with  $\approx400\,$s time delay after the black hysteresis. Between these two measurements the time-dependence of the conductance was recorded at the signed displacement values (green dots) at $V=150\,$mV bias voltage, as demonstrated in panels b)-d). b) Outside of the hysteresis at large displacement the LowG configuration is stable and no switching was observed. c) At somewhat smaller displacement a bistable switching occurs and the LowG configuration is preferred. d) At even smaller displacement the switching is reversed, and most of the time is spent in the HighG configuration.
    e) The distribution of the times spent in the on (red) and off (blue) states between subsequent switching events at the displacement position c). The histogram bins are equally spaced on the logarithmic scale. The dashed lines demonstrate the average $\tau_\text{on}$ and $\tau_\text{off}$ times, whereas the solid lines represent the best fitting exponential probability density functions. f) Similar distributions at the displacement position d).}
    \label{fig:fig4}
\end{figure}

Next, we correlate the bistable electric switching with the displacement of the electrodes. During this measurement, we identify the displacement interval of the hysteretic conductance switching at low bias ($V=30\;$mV, see black line in Fig. \ref{fig:fig4}a), and afterwards we stop at distinct displacement values (green signs in Fig. \ref{fig:fig4}a) to record the temporal evolution of the junction conductance under an elevated bias voltage ($V=150\,$mV). The corresponding representative segments of these time-traces are shown in Figs.~\ref{fig:fig4}b,c,d. Figs.~\ref{fig:fig4}e,f demonstrate the distribution of the times spent in the  HighG ($\tau_{\text{on}}$, red) and LowG ($\tau_{\text{off}}$, blue) conductance states between two subsequent switching events for the entire length of the time traces ($22\,$s) at the displacement points c) and d), where a telegraph-like conductance fluctuation was observed. By recording the low bias conductance hysteresis at the end of the measurement again (grey line in Fig. \ref{fig:fig4}a) we estimate the stability of the junction to be on the scale of $<40\,$pm on the timescale of the entire measurement ($\approx400\,$s). This displacement inaccuracy is $\approx 15\%$ of the Au-Au interatomic distance in a gold chain. Based on these measurements, and further similar measurements we find the following clear tendencies: (i) the bistable electrical switching is focused around the displacement range of the low-bias conductance hysteresis (see Figs. \ref{fig:fig4}c,d). (ii) Within this range the average times spent in the two states cross, i.e. at smaller displacement the HighG state, whereas at larger displacement the LowG state is dominant (Figs. \ref{fig:fig4}e,f). (iii) further away from the hysteresis, the electrical switching vanishes (Fig. \ref{fig:fig4}b), and the corresponding state is stable even under an elevated voltage.

\begin{figure}
    \centering
    \includegraphics[width=0.9\columnwidth]{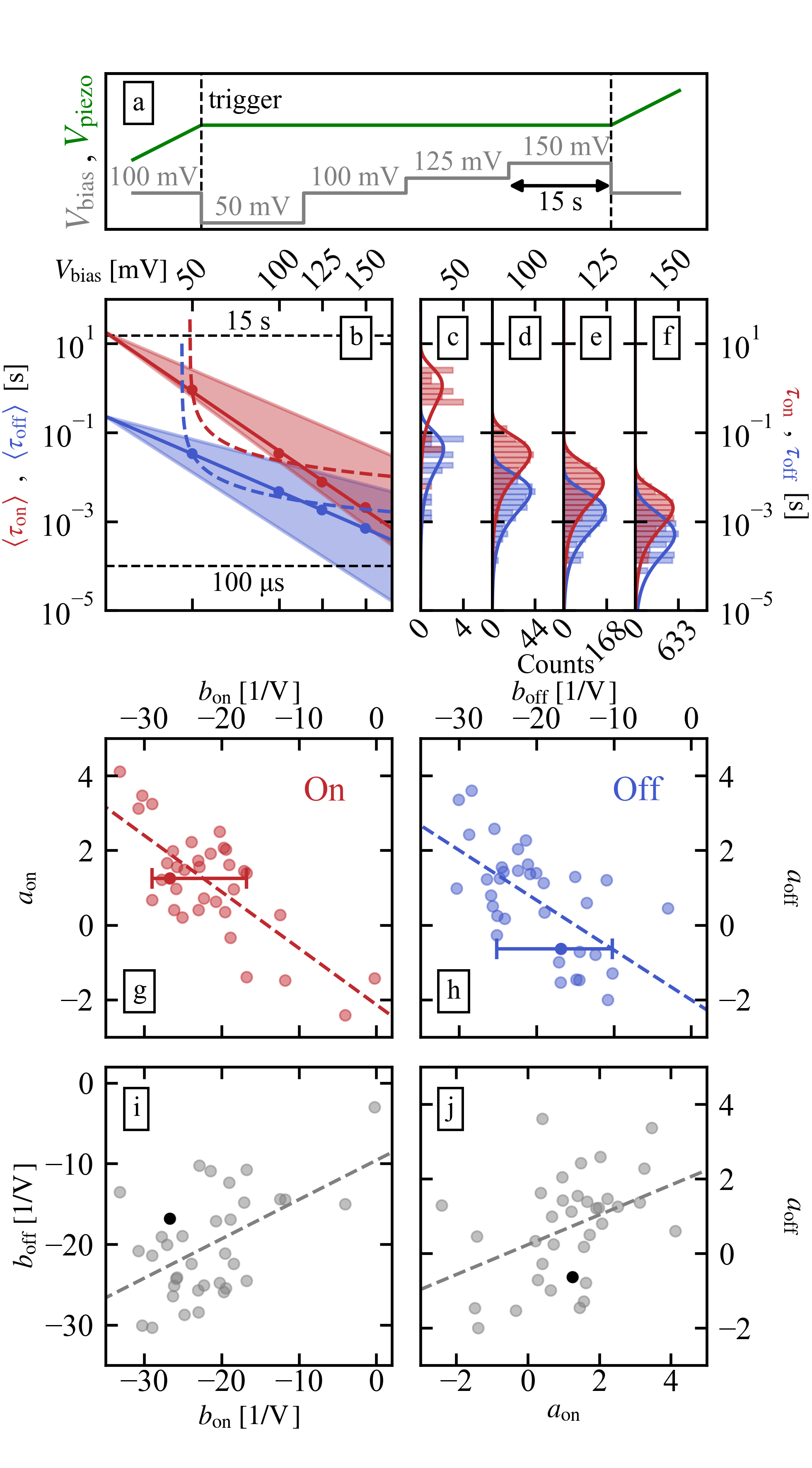}
    \caption{Statistical investigation of the voltage dependent switching times using an automated measurement setup. \textbf{a)} The scheme of the measurements (see text). \textbf{b)-f)} The voltage dependence of the switching times in one of the measured single-molecule junctions. c)-f) The distribution of the times spent in the on/off states between subsequent switching events at various bias values together with the best fitting exponential probability density functions. The histogram bins are equally spaced on the logarithmic scale. b) The voltage dependence of the average switching times deduced from the distributions in panels c)-f). The solid lines represent the fitting curves considering an exponential speedup of the switching time upon linearly increasing voltage (see text). The black horizontal dashed lines illustrate the limiting time-scales, i.e. the length of the measurement, and the bandwidth limit of the measurement. The red and blue dashed lines represent the best fitting curves considering a simple two-level system model with a well-defined excitation energy and a voltage independent coupling strength, i.e. $\tau=A/(eV-E)$. \textbf{g)-j)} the dependence of the $a_\text{on/off}$ intercept and $b_\text{on/off}$ slope values on each other for all the measured data together with the best fitting linear functions: g): $a_\text{on}= -2.135 - 0.151 \cdot b_\text{on}$; h): $a_\text{off} = -2.006 - 0.134 \cdot b_\text{off}$; i): $b_\text{off}= -9.588 + 0.487 \cdot b_\text{on}$; j): $a_\text{off}= 0.23 + 0.400 \cdot a_\text{off}$.}
    \label{fig:fig5}
\end{figure}

To resolve the temporal dynamics of the switching, we have performed a statistical amount of time-resolved measurements at various driving voltages. This was achieved by an automated measurement setup, where a real-time FPGA controller monitored the conductance and stopped the junction elongation once the predefined trigger conductance of $G=0.005\,$G$_0$ was reached. At this point, the displacement (i.e. the voltage on the piezo actuator) was fixed, and a voltage staircase was applied on the junction (see the scheme of the measurement in Fig.~\ref{fig:fig5}a). Around $8\%$ of the 8000 automated measurements yielded molecular contacts that were stable throughout the voltage sweep. Note, that this number is partly related to the low molecular pick-up rate in the low-temperature measurements. \cite{magyarkuti2020} Out of these stable molecular contacts $\approx5\%$ matched the displacement range, where the voltage-dependent bistable switching was observable. These 35 independent traces are used to investigate the voltage-dependence of the bistable switching in Figs.~\ref{fig:fig5}b-j. One of these measurements is exemplified in Figs.~\ref{fig:fig5}c-f showing the distribution of the times spent in the HighG and LowG conductance states at $50\,$mV, $100\,$mV, $125\,$mV and  $150\,$mV driving voltages respectively.  We find, that the distribution of the $\tau_{\text{on}}$ and $\tau_{\text{off}}$ times are well fitted by an exponential probability density function (see the solid lines in Figs.~\ref{fig:fig5}c-f and in Figs.~\ref{fig:fig4}e-f as well). The exponential distribution implies stochastic switching events, independent of the elapsed time since the last switching event. The voltage dependence of the average times $\left\langle \tau_{\text{on}}\right\rangle$ and $\left\langle \tau_{\text{off}}\right\rangle$
are plotted in Fig.~\ref{fig:fig5}b by red and blue dots, respectively. The solid lines show that both voltage dependencies are well fitted with a linear function once a logarithmic timescale is applied, i.e. $\log \left\langle\tau_\text{on/off}/\tau_0\right\rangle=a_\text{on/off}+b_\text{on/off}\cdot V$, respectively, where we use the reference time $\tau_0=1\,$s to make the argument of the logarithm dimensionless. This means, that the switching exponentially speeds up upon linearly increasing the voltage. Figs. \ref{fig:fig5}g-j show the correlations between the $a_\text{on/off}$ intercept and $b_\text{on/off}$ slope values. Despite the scattering of the data, clear tendencies are observable: (i) for a certain (on/off) state a larger intercept is accompanied by a larger slope (see Figs. \ref{fig:fig5}g,h); (ii) a larger slope/intercept value for a certain state is accompanied by a larger slope/intercept for the other state as well (see Figs.~\ref{fig:fig5}i,j). The general tendencies are also illustrated by the dashed lines, i.e. the linear functions best fitting the data. The bright red/blue dots in Figs.~\ref{fig:fig5}g,h, and the black dots in  \ref{fig:fig5}i,j correspond to the measurement data in Fig.~\ref{fig:fig5}b. The horizontal red/blue error bars in Figs.~\ref{fig:fig5}g,h illustrate the statistical scattering of the slope values for the rest of the data points with similar intercept. This is also illustrated by the shaded areas in Fig.~\ref{fig:fig5}b, where the intercepts are fixed at the actual values of the corresponding data set, while the opening of the shaded areas reflect the scattering of the slope values according to the error bars in Figs.~\ref{fig:fig5}g,h.
These figures demonstrate a clear tendency: the on and off switching times speed up \emph{simultaneously}, and the dominance of a certain state is usually preserved in the investigated voltage range. Note that this is in sharp contrast to the mechanical manipulation, which clearly reverses the dominance of the states, as demonstrated in Figs.~\ref{fig:fig4}e,f.

\begin{figure}
    \centering
    \includegraphics[width=0.8\columnwidth]{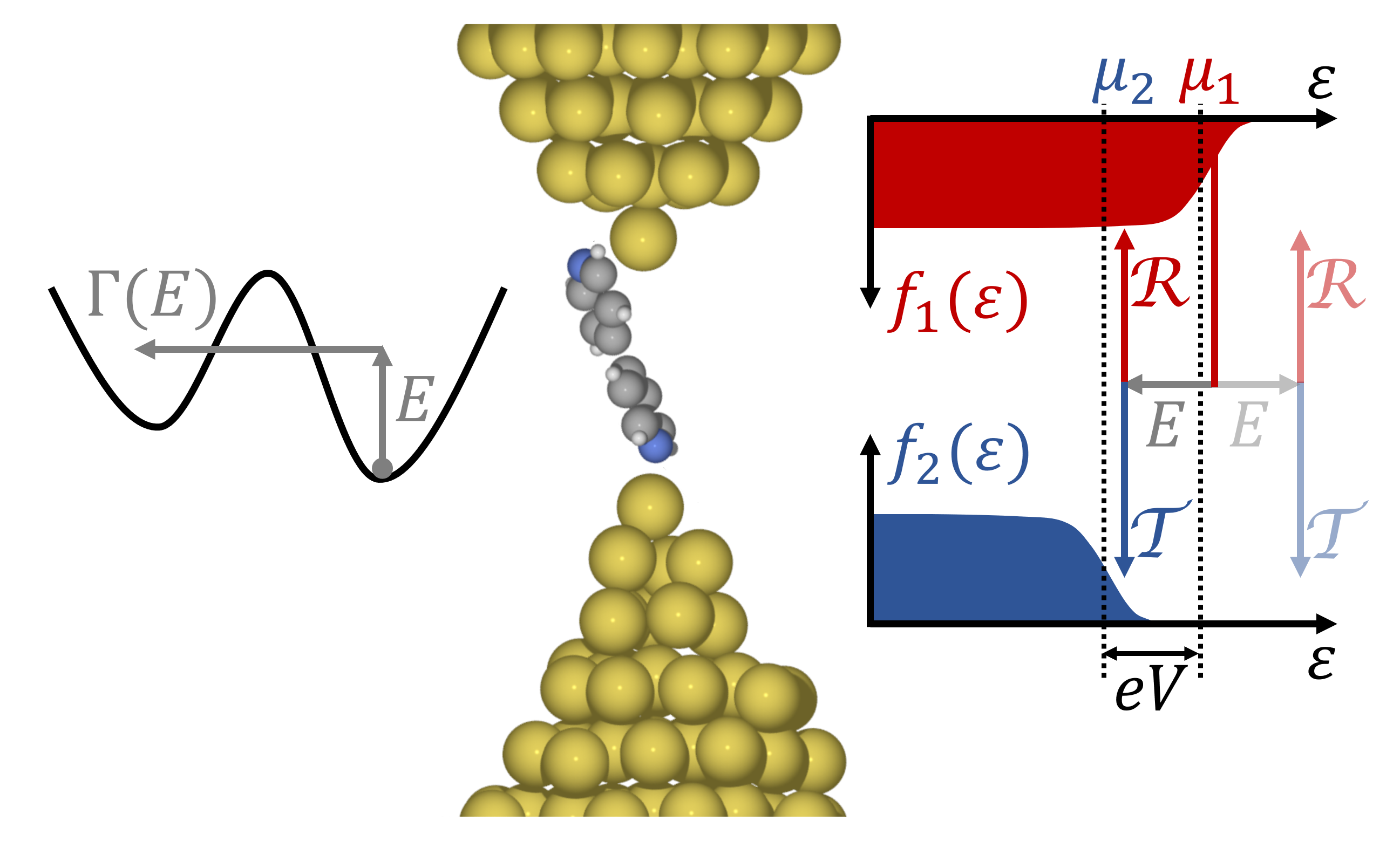}
    \caption{Illustration of the model considerations. The probability of the switching event includes two components: (i) the probability that an electron from an occupied state releases $E$ energy ending up in an unoccupied state (right cartoon), and (ii) the $\Gamma(E)$ probability that the molecule can switch to its other state after gaining $E$ energy (left cartoon). Note, that a reflected electron does not find unoccupied final states at $E\gg kT$ energy release, i.e. this process requires the transmission through the molecule with $\mathscr{T}\approx 10^{-3}$ probability. In a vibrational pumping model one also has to consider the process, where the electrons absorb $E$ energy from the molecular vibrations (see the semitransparent arrows in the right cartoon). In this case a reflected electron also finds unoccupied final states, and therefore the $\mathscr{R}/\mathscr{T}\approx10^{3}$ ratio makes the cooling of the vibrational mode much more efficient than the pumping.}
    \label{fig:fig6}
\end{figure}

Finally, we discuss our results on the bistable resistance switching in terms of simple model considerations (See Fig.~\ref{fig:fig6}). According to the force measurements and computer simulations in Refs.\citenum{force6,Aradhya2014}, the energy barrier between the two states of the Au-BP-Au single-molecule junction is on the order of $\approx 500-700\,$meV, which is somewhat larger, but the same order of magnitude as the voltage scale of the switching in our measurements. We consider that in a single scattering event, the electrons release an energy $E$ to the molecule. From the electrons' point of view the probability of such event, within a unit time interval, scales with $\mathscr{T}\cdot\int f_1(\varepsilon)\cdot (1-f_2(\varepsilon-E))$, where $f_1$ and $f_2$ are the Fermi functions of the two electrodes with a chemical potential difference $\mu_1-\mu_2=eV$, and $\mathscr{T}$ is the transmission probability through the molecule. Applying the $\int f_0(\varepsilon)(1-f_0(\varepsilon-a))=(a/2)\cdot(\coth(a/2kT)-1)$ identity for the $f_0$ Fermi function the above integral simplifies to  $\mathscr{T}\cdot(E-eV)\cdot\exp{\left(-(E-eV)/kT\right)}$ at $eV<E$. Such a temperature-activated process is unrealistic at $4.2\,$K, as the energy scale in the numerator of the exponent is almost $3$ orders of magnitude larger than the thermal energy in the denominator. Furthermore, this scaling would provide a slope of $b=-e\cdot\log(\mathrm{e})/kT\approx-10^3\,$V$^{-1}$, whereas the largest measured slopes are restricted to $b\approx-30\,$V$^{-1}$ (see Fig.~\ref{fig:fig5}i). 

We can also analyze the processes, where the electrons release a smaller energy to the molecule than the applied voltage, i.e. $eV>E$. In this case the above Fermi-function integral simplifies to $\mathscr{T}\cdot(eV-E)$. \cite{NDC} In a single two-level system model with a well-defined excitation energy $E$ and a voltage-independent interaction matrix element between the electrons and the molecule the inverse of this function describes the voltage scaling of the switching times: $\tau\sim (eV-E)^{-1}$.\cite{Fredi_TLS,NDC} The red and blue dashed lines in Fig. \ref{fig:fig5}a illustrate the best fits with this voltage scaling, which is clearly inconsistent with the measured data. Even if the divergence at $eV=E$ is blurred due to a broader energy distribution,  the high bias $1/V$ scaling of this simple model is much shallower than the measured voltage dependence of the data. Note, that a $1/V$ dependence would yield a factor of $3$ speedup between $50$ and $150\;$mV driving in contrast to the observed $>2$ orders of magnitude speedup. 

So far we have considered a simple model, where the energy release from the electrons kicks the molecule to its other state within a single event. Alternatively, the electrons may release smaller $E=\hbar\omega$ vibrational quanta to the molecule, and the transition happens once this vibrational degree of freedom is pumped \cite{Todorov1998,pumping02} to a high enough threshold energy. This process, however, is damped due to the small, $\mathscr{T}\approx 10^{-3}$ transmission probability of the single-molecule junction. Note, that the electrons can only release energy to the molecule once they are transmitted from the first electrode with higher chemical potential to the second one with lower chemical potential, i.e. the probability of this event scales with $\mathscr{T}$. On the other hand, the electrons can absorb energy from the vibrational mode of the molecule also during a reflection event, for which the probability scales with $\mathscr{R}=1-\mathscr{T}$. This $\mathscr{R}/\mathscr{T}\approx10^3$ imbalance factor between the emission/absorption probabilities makes the vibrational pumping model unrealistic. 

Finally, in our system we can also rule out the model describing the voltage-induced switching relying on the azobenzene side group of a terphenyl aromatic chain between graphene point contacts. \cite{Meng2019} In the latter work the low bias excited state became dominant at high bias due to the voltage-induced inversion of the minima in the double-well potential, which was reflected by the clear crossing in the voltage dependence of the switching times. In our system the dominance of the states is manipulated by the displacement, but at a fixed electrode separation the switching of the two configurations speed up simultaneously, and the initially dominant state remains dominant throughout the studied voltage train. 

To be able to describe the simultaneous exponential decay of the switching times with linearly increasing voltage, we consider a strong energy dependence of the molecule's transition probability $\Gamma(E)$ (see left illustration in Fig. \ref{fig:fig6}). This may relate to an energy-dependent tunnelling probability between two conformational states, as implied on the figure, but several other processes, like the reduction of the barrier with the applied voltage can be considered. A more detailed understanding of the precise microscopic processes would require first principle model calculations.

\section{Conclusions}

In conclusion we have demonstrated well-controlled voltage-induced probabilistic conductance switching in cryogenic temperature Au-BP-Au single-molecule junctions. This behavior resembles probabilistic bit devices with two tunable parameters: the relative dominance of the two conductance states is manipulated by the electrode displacement, whereas the voltage manipulation induces an exponential speedup of both switching times.    

We have compared our experimental results with possible model considerations, and argued against a \emph{simple two-level system model}, a \emph{temperature activated switching model}, a \emph{vibrational pumping model} and a \emph{double-well potential inversion model}. Finally we proposed a heuristic model relying on the exponential energy dependence of the molecule's transition probability. 

\section*{Conflicts of interest}
There are no conflicts to declare.

\section*{Acknowledgements}
This work was supported by the BME-Nanonotechnology FIKP grant of EMMI (BME FIKP-NAT) and the NKFI K119797 grant.

\section*{Author contributions}
The measurements were performed and analyzed by G. Mezei and Z. Balogh. The real-time-controlled measurement system was developed and programmed by A. Magyarkuti. The project was supervised by A. Halbritter. All authors contributed to the discussion and interpretation of the results. The manuscript was written by A. Halbritter, G. Mezei and Z. Balogh with inputs from all authors.

\bibliography{manuscript_final_arxiv} 
\bibliographystyle{manuscript_final_arxiv}

\end{document}